\documentclass[3p, twocolumn, 10pt]{elsarticle}
\usepackage{floatrow}
\usepackage{booktabs,caption}
\usepackage[flushleft]{threeparttable}
\usepackage{multirow}
\usepackage{makecell}
\setcellgapes{3pt}

\usepackage{amssymb}
\usepackage{amsmath}
\usepackage{xcolor}
\newcommand{\boldgreek}[1]{{\mbox{\boldmath$ {#1} $}}}
\newcommand{\+}{$^+$}
\newcommand{\SiO}[1]{$^{#1}$Si$^{16}$O\+}

\newcommand{\ket}[1]{{\left| {#1} \right\rangle}}

\begin{document}

\begin{frontmatter}



\title{High-resolution laser-induced fluorescence spectroscopy of \SiO{28} and \SiO{29} in a cryogenic buffer-gas cell} 


\author[inst1]{Guo-Zhu Zhu}

\affiliation[inst1]{organization={Department of Physics and Astronomy},
            addressline={University of California, Los Angeles}, 
            city={Los Angeles},
            postcode={90095}, 
            state={CA},
            country={USA}}

\author[inst1]{Guanming Lao}

\author[inst1]{Clayton Ho}

\author[inst1,inst2,inst3]{Wesley C. Campbell}

\author[inst1,inst2,inst3]{and Eric R. Hudson}

\affiliation[inst2]{organization={Center for Quantum Science and Engineering},
            addressline={University of California, Los Angeles}, 
            city={Los Angeles},
            postcode={90095}, 
            state={CA},
            country={USA}}

\affiliation[inst3]{organization={Challenge Institute for Quantum Computation},
            addressline={University of California, Los Angeles}, 
            city={Los Angeles},
            postcode={90095}, 
            state={CA},
            country={USA}}

\begin{abstract}
The electronic, laser-induced fluorescence spectrum of the $B^2\Sigma^+\leftarrow X^2\Sigma^+$ transition in \SiO{28} and \SiO{29} has been recorded in a cryogenic buffer gas cell at $\approx 100$~K.
Molecular constants are extracted for both \SiO{28} and \SiO{29}, including the Fermi contact hyperfine constant for both the $B$ and $X$ states of \SiO{29}, and used in a discussion of the suitability of SiO$^+$ in future quantum information experiments. 
\end{abstract}




\end{frontmatter}



\section{Introduction}

With the successful application of laser cooling and trapping of atoms to quantum information science \cite{saffman2010quantum, ludlow2015optical}, it is natural to consider extending these techniques to molecular systems \cite{mccarron2018laser,tarbutt2019laser}.
Molecules possessing rich internal structure provide new capabilities in a wide range of fields, including new platforms for quantum simulation and computation \cite{blackmore2018ultracold, yu2019scalable}, tests of fundamental physics \cite{augenbraun2020laser, hutzler2020polyatomic}, and ultracold chemistry and collisions \cite{balakrishnan2016perspective, liu2021precision}. 
As proposed by Di~Rosa \cite{di2004laser}, one of the primary requirements of direct laser cooling of molecules is that the molecules possess diagonal Franck-Condon factors (FCFs), which suppresses spontaneous decays that change the molecule vibrational state and thereby enable closed optical cycling -- i.e., spontaneously scattering many photons following optical excitation in a repeated cycle. 

While several neutral molecules have now been laser cooled, such as SrF \cite{shuman2010laser}, CaF \cite{truppe2017molecules, anderegg2017radio}, YbF \cite{lim2018laser}, YO \cite{collopy20183d}, SrOH \cite{kozyryev2017sisyphus}, and CaOH \cite{baum20201d}, YbOH \cite{augenbraun2020laser} and CaOCH$_3$ \cite{mitra2020direct}, comparatively little experimental work has been carried out on molecular ions. 
Recently, we undertook a systematic search for diatomic molecular ions suitable for optical cycling~\cite{ivanov2020search} and found, in addition to four candidates already proposed by other workers, SiO\+ \cite{nguyen2011prospects,li2019laser}, BH\+ \cite{nguyen2011challenges, zhang2017low}, AlH\+ \cite{lien2014broadband} and AlCl\+ \cite{kang2017ab}, only three additional potential candidates: BO\+, PN\+, and YF\+.

Of these candidate diatomic molecular ions, \SiO{28}, pioneered by the Odom group~\cite{nguyen2011challenges}, currently appears the most promising. 
The vibrationless branching ratio of the $B^2\Sigma^+ \rightarrow X^2\Sigma^+$ transition has been measured by laser-induced fluorescence (LIF) of a supersonic \SiO{28} beam and found to be  $0.970^{+0.007}_{-0.025}$~\cite{stollenwerk2017electronic}. 
Rotational cooling of \SiO{28} has recently been achieved by selectively driving the P-branch transition using a broadband laser~\cite{stollenwerk2020cooling} and optical pumping of trapped \SiO{28} to a super-rotor state has also been demonstrated~\cite{antonov2021rotor}.  

When looking forward to future quantum applications of SiO\+, a natural question arises: \emph{How is a qubit best hosted in SiO\+?}
Typically, it is proposed to encode quantum information in the rotational degree of freedom of polar molecules.
However, as discussed in Ref.~\cite{hudson2018dipolar}, for molecular ions held in an ion trap, thermal motion leads to rotational decoherence via the Stark shift caused by the electric fields of the trap and the other ion monopoles. 
Alternatively, a qubit could be hosted in the ground-state Zeeman structure to mitigate Stark-induced decoherence, but such qubits are complicated by the requirements for magnetic field control~\cite{hudson2018dipolar, Ruster2016LongLived}.
An attractive alternative is to use an isotopologue of \SiO{28} that has non-zero hyperfine structure to realize a hyperfine qubit. 
Conveniently, the naturally occuring isotope ${}^{29}\mathrm{Si}$ has a nuclear spin of $I = 1/2$, leading to a ground-state characterized by total angular momentum $F = 0$ and $F = 1$ states. 
Such structure provides a so-called zero-field clock-state qubit in the two $M=0$ Zeeman sublevels, which in atomic ion quantum computing has been shown to have attractive decoherence properties~\cite{Christensen2020high, roman2020coherent}.
It also provides a frequency resolved qubit state ($|F=0\rangle$), which can be prepared and detected with frequency selective optical pumping. 

While \SiO{29} will share the diagonal FCFs of \SiO{28}, to the best of our knowledge there has not yet been a spectroscopic study of gas-phase \SiO{29}. 
Here we report spectroscopy of the (0,0) band of $B^2\Sigma^+ \leftarrow X^2\Sigma^+$ of  \SiO{29}.
We find that \SiO{29} is best described with Hund's case (b)$_{\beta S}$ coupling scheme as the hyperfine interaction exceeds the spin-rotation coupling. 
We also report spectroscopy of the  (0,0)  and  (1,0) bands of $B^2\Sigma^+ \leftarrow X^2\Sigma^+$ of \SiO{28} and find spin-rotation constants for both the $X^2\Sigma^+$ and $B^2\Sigma^+$ states that in some casese are incompatible with previously recommended values \cite{ghosh1979confirmation, zhang1993new, cameron1995fast1, cameron1995fast2}. 

\section{The SiO\+ molecule}
In addition to its interest for quantum information, SiO\+ plays an important role in silicon chemistry of diffuse interstellar clouds \cite{turner1977Sichem, herbst1989SiChem, langer1990SiChem} and circumstellar regions \cite{clegg1983circumstellar}, and
its electronic structure and spectroscopic properties have been extensively investigated. 
In 1940, Pankhurst \cite{pankhurst1940band} first observed a band of SiO\+ at around 3840 \r{A} from a heavy-current hydrogen discharge tube with a quartz constriction, but incorrectly ascribed it to SiO$_2$. 
The SiO\+ band was later identified by Woods \cite{woods1943silicon} in 1943 as the $B^2\Sigma^+-X^2\Sigma^+$ transition and further confirmed by Ghosh et al. \cite{ghosh1979confirmation} using the isotope shift measurement of \SiO{28} and $^{28}$Si$^{18}$O\+. 
The molecular constants have also been obtained for both \SiO{28} and $^{28}$Si$^{18}$O\+. 
Using the vacuum UV photoelectron spectrosocpy of SiO ($X^1\Sigma^+$), Colbourn and coworkers \cite{colbourn1978vacuum} have estimated the spectroscopic constants for three ionic states of \SiO{28}:  $X^2\Sigma^+$,  $A^2\Pi$, and $B^2\Sigma^+$. 
Later, Rosner and coworkers \cite{zhang1993new, cameron1995fast1, cameron1995fast2, rosner1998study} measured a number of high-resolution bands of $A^2\Pi-X^2\Sigma^+$ and $B^2\Sigma^+-X^2\Sigma^+$ transitions of \SiO{28} from LIF studies of mass-selected fast ion beams. By fitting the entire data with a Hamiltonian model, they have obtained more precise rotational and fine-structure constants for all three states. 
Along with the experimental studies, many theoretical calculations have been performed to investigate the potential energy curves and spectroscopic constants of the low-lying electronic states of SiO\+ \cite{chong1977vibrational, werner1982ab, cai1998ab, cai1998internally, cai1999ab, Chattopadhyaya2003ElectronicSO, shi2012mrci, li2019explicitly, qin2020transition}. 

\section{Hamiltonians for the X$^2\Sigma$\+ and B$^2\Sigma$\+ states of SiO\+}
The effective 
Hamiltonian for a molecule in a $^2\Sigma^+$ state is:
\begin{equation}
\label{Heff}
    H_\mathrm{eff}=H_\mathrm{el}+H_\mathrm{vib}+H_\mathrm{rot}+H_\mathrm{hfs},
\end{equation}
where the four contributions correspond to the electronic, vibrational, rotational and hyperfine-structure Hamiltonian, respectively. 
The energy of electronic state is conventionally denoted as $T_e$, and the vibrational Hamiltonian is
\begin{equation}
    H_{vib}/h = \omega_e\left(v+\frac{1}{2}\right)-\omega_e x_e\left(v+\frac{1}{2}\right)^2,
\end{equation}
where $\omega_e$ and $\omega_e x_e$ are vibrational constants, and $v$ is the vibrational quantum number. 
The vibronic part of the Hamiltonian can then be rearranged as:
\begin{equation}
    H_{el}/h+H_{vib}/h=T_0+[\omega_e-\omega_e x_e\left(v+1\right)]v,
\end{equation}
where the electronic energy $T_e$ and the vibrational ground state energy 
are absorbed into  $T_0$.

The rotational Hamiltonian reads
\begin{equation}
\label{Hrot}
    H_{rot}/h =B_v\boldsymbol{N}^2-D_v\boldsymbol{N}^4+\gamma T^1(\boldsymbol{N})\cdot T^1(\boldsymbol{S}), 
\end{equation}
where $\boldsymbol{N}$ is the rotational angular momentum of the molecule about its center of mass, $\boldsymbol{S}$ is the electron spin, $B_v$ and $D_v$ are molecular rotational constants of vibrational state $|v\rangle$, and $\gamma$ is the spin-rotation coupling constant. 
Here all rotational constants include the contribution from rotation-vibration coupling. 
The third term in Eq. (\ref{Hrot}) is the electron spin-rotation interaction, represented as the scalar product of two spherical tensors, defined via $T^k(\mathbf{A})\cdot T^k(\mathbf{B}) \equiv \sum_p (-1)^p T^k_p(\mathbf{A})T^k_{-p}(\mathbf{B}) $.  
\begin{table*}[t]
    \centering
    \begin{tabular}{c c c c c}
    \hline
    \hline
     & $G'=1, F'=N+1$ & $G'=1, F'=N$ & $G'=0, F'=N$ & $G'=0, F'=N-1$\\
    \hline
     $G=1, F=N+1$    & $\frac{\gamma N}{2}+\frac{b_F}{4}-\frac{Nt}{4N+6}$ &0&0&0\\
     $G=1, F=N$ &0&$-\frac{\gamma}{2}+\frac{b_F}{4}+\frac{t}{2}$&$\frac{\gamma}{2}\sqrt{N(N+1)}$&0\\
     $G=0, F=N$& 0 &$\frac{\gamma}{2}\sqrt{N(N+1)}$& $-\frac{3b_F}{4}$&0\\
     $G=1, F=N-1$ &0&0&0&$-\frac{\gamma(N+1)}{2}+\frac{b_F}{4}-\frac{t(N+1)}{4N-2}$\\
     \hline
     \hline
     \multicolumn{5}{l}{\begin{tabular}{l}
     Note: For $N=0$, there are only two non-zero elements: $G=G'=1, F=F'=1$ and $G=G'=0,$\\
      $F=F'=0$.  The elements off-diagonal in $N$ are neglected because of the large energy gap between the\\  rotational levels.\end{tabular}}
    \end{tabular}
    \caption{Matrix elements of the hyperfine and spin-rotation interactions. 
    }
    \label{HMET}
\end{table*}

The hyperfine Hamiltonian, nonzero for \SiO{29}, is
\begin{equation}
\label{Hhfs}
\begin{split}
    H_{hfs}/h =&b_FT^1(\boldsymbol{I})\cdot T^1(\boldsymbol{S})+\sqrt{6}g_S\mu_Bg_N\mu_N\\
    &\times (\mu_0/4\pi)T^2(\boldsymbol{S},\boldsymbol{I})\cdot T^2(\boldsymbol{C}),
\end{split}
\end{equation}
where $T^2$ indicates a spherical tensor of rank 2, $T^2(\boldsymbol{C})$ represents a normalized spherical harmonic,  $T^2(\boldsymbol{S, I})$ describes the dipolar coupling between spins, and the rest of symbols have their standard meanings \cite{brown2003rotational}. 
The first and the second terms represent the Fermi contact and dipolar interactions, respectively. 
The matrix element of the dipolar interaction term is parameterized by  \textit{t}, which is equal to the perhaps more common dipolar hyperfine constant $t_0$ in Hund's case (b)$_{\beta J}$, and is given as \cite{brown2003rotational}
\begin{equation}
\begin{split}
    t&=t_0=g_S\mu_Bg_N\mu_N(\mu_0/4\pi)\langle\eta|T^2_0(\boldsymbol{C})|\eta\rangle,
\end{split}
\end{equation}
where $|\eta\rangle$ is the electronic wave function. 

For the $X^2\Sigma$ states of \SiO{29}, the Fermi contact interaction is much stronger than the spin-rotational coupling \cite{knight1985generation, ghosh1979confirmation} (i.e., $b_F\gg\gamma$). 
A $^2\Sigma$ molecule with strong spin-spin interaction is conveniently described by Hund's case (b)$_{\beta S}$, 
in which the spins $S$ and $I$ couple to form a resultant $G$, and $G$ then couples with $N$ to form total angular momentum $F$. 

With this basis choice, the matrix elements of the spin-rotation and hyperfine interactions are calculated and listed in Table~\ref{HMET}. 
As can be seen in this table, the spin-rotation coupling term in $H_{rot}$ mixes states with different $G$ but the same $F$. 
We use this basis for both the $X^2\Sigma$\+ and $B^2\Sigma$\+ states of SiO\+ and
the molecular constants of each state are distinguished by superscripts ($X$) and ($B$) in what follows.

The line strength of a transition between these states, i.e. $|\eta_X, N'', G'', F''\rangle$ and $|\eta_B, N', G', F'\rangle$, is \cite{western2017pgopher}:
\begin{equation}
\begin{split}
    &S_{N'G'F'N''G''F''} 
    =(2F'+1)(2F''+1)(2N'+1)\\
    &\times(2N''+1)\delta_{G'G''}T^{(1)}_{q=0}(\boldgreek{\mu})^2\begin{Bmatrix} N' & F'& G' \\ F''& N'' &1\end{Bmatrix}^2,
\end{split}
\end{equation}
where $\delta_{G'G''}$ is the Kronecker delta function, $T^{(1)}_{q=0}(\boldgreek{\mu})$ is the dipole transition moment between the electronic states $|\eta_B\rangle$ and $|\eta_X\rangle$ in the body-fixed frame, and $\left\{:::\right\}$ is the Wigner-6j symbol. 
The Kronecker delta describes the useful fact when interpreting spectra that the electron and nuclear spin are unchanged by the electric dipole transition. 
Use of the line strength to predict the relative strength of the observed transitions is helpful in assigning the observed lines and more details are given in the data analysis session.


\section{Apparatus}
In this experiment, SiO\+ ions were produced by direct reaction of Si\+ and O$_2$ \cite{matsuo1997formation} in a cryogenic cell at a temperature $\approx 90$~K, as schematically shown in Fig.\ 1. 
Si\+ ions were generated by ablating a Si target using a pulsed Nd:YAG laser at 1064 nm focused onto the target with a spot size $\approx 1$~mm and a pulse energy $\approx 5$ mJ. 
A silicon wafer (Sigma Aldrich, natural isotopic abundance) was used to produce $^{28}$Si\+ and pressed target of $^{29}$Si metalloid powder (Buyisotope, isotopic enrichment $>$99.2\%) was used for $^{29}$Si\+. 
The produced Si\+ ions then reacted with ultrahigh purity O$_2$ gas (Praxair, 99.993\%), which was introduced into the cell at a flow rate $\approx 7$ standard cubic centimeters per minute (sccm) resuling in a gas density $\approx 10^{15}-10^{16} ~\mathrm{cm}^{-3}$, to form SiO\+ ions.
\begin{figure}[h]
    \centering
    \includegraphics[scale=0.9]{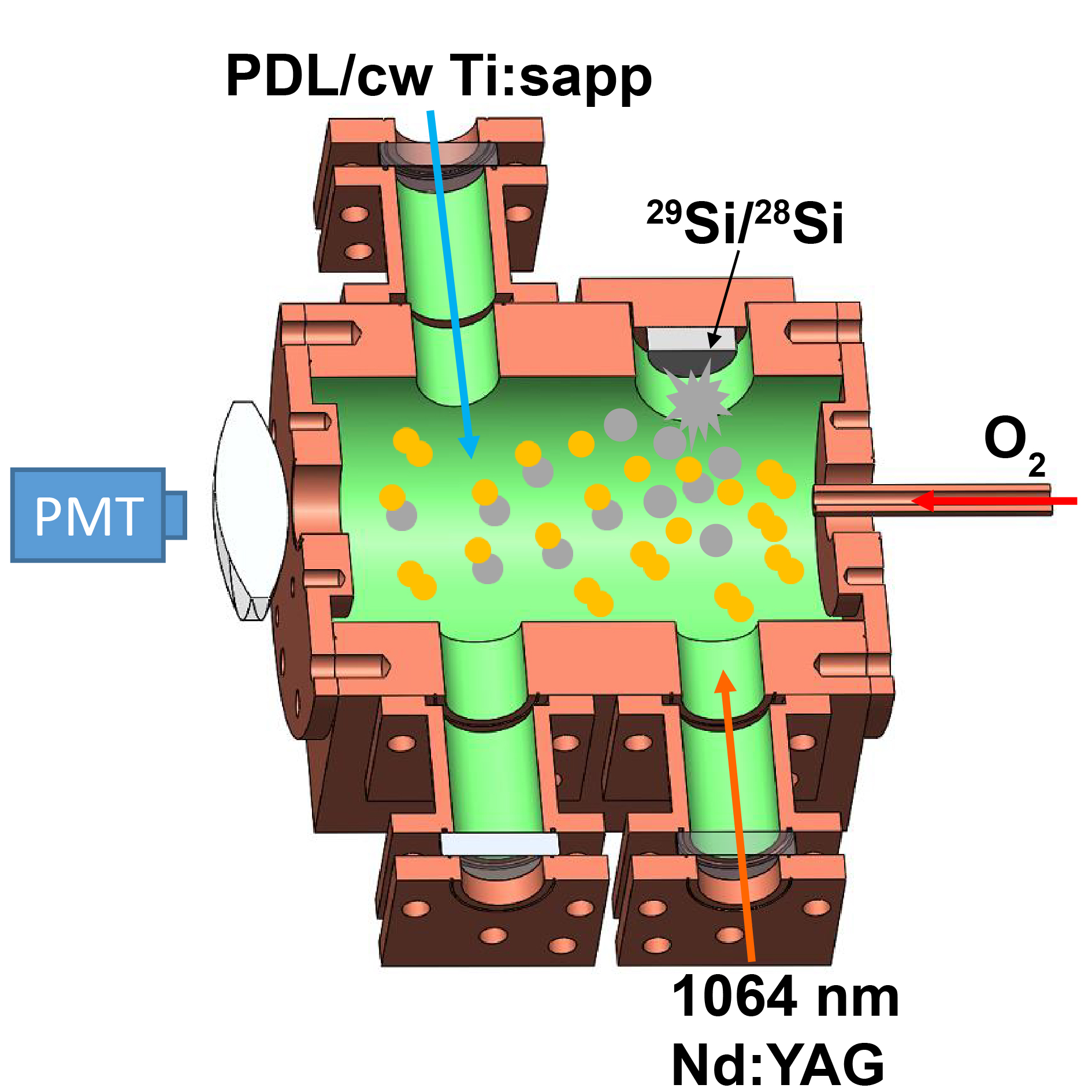}
    \caption{Schematic of the cryogenic cell with production of SiO\+ and LIF indicated.}
    \label{PDLexp}
\end{figure}

The \SiO{28} ions were first probed at 368~nm and 384~nm, generated by frequency doubling of a pulsed dye laser with LDS 751 laser dye (LiopStar-E dye laser, linewidth 0.04~cm$^{-1}$ at 620~nm), to cover the (0,0) and (1,0) bands of $B^2\Sigma^+ \leftarrow X^2\Sigma^+$, respectively. To avoid scattered light from the ablation plume, the dye laser illuminated the ions roughly 120~$\mu$s after the ablation pulse.  
A Coherent WaveMaster was used to calibrate the absolute wavelength of the dye laser.
The resulting fluorescence photons were imaged onto a PMT via a lens system and counted (Stanford Research Systems, SR430). 
Both bands were scanned at a step size of 2 GHz near the resonance and 4 GHz when away from resonance. 
Each data point 
is the sum of 500 laser pulses at a 10~Hz repetition rate and repeated between two and four times.

Following the dye laser survey spectroscopy, higher resolution spectroscopy was performed on the (0,0) bands of \SiO{28} and \SiO{29} using a doubled, continuous wave Ti:sapphire laser (M Squared Lasers). 
This laser has an effective linewidth on the order of 10~MHz 
and was scanned with a step size between 50~MHz and 200~MHz. The laser frequencies were measured and recorded using a High Finesse WS-U wavemeter.
The resulting LIF photons were counted over a 5~$\mu$s window that was delayed by 120~$\mu$s from the ablation pulse and averaged for 300 ablation pulses.  
Each measurement was repeated between two and four times.

\section{Results and discussion}
\subsection{Pulsed dye laser spectroscopy}
\begin{figure*}[h!]
    \centering
    \includegraphics[width = 0.98\textwidth]{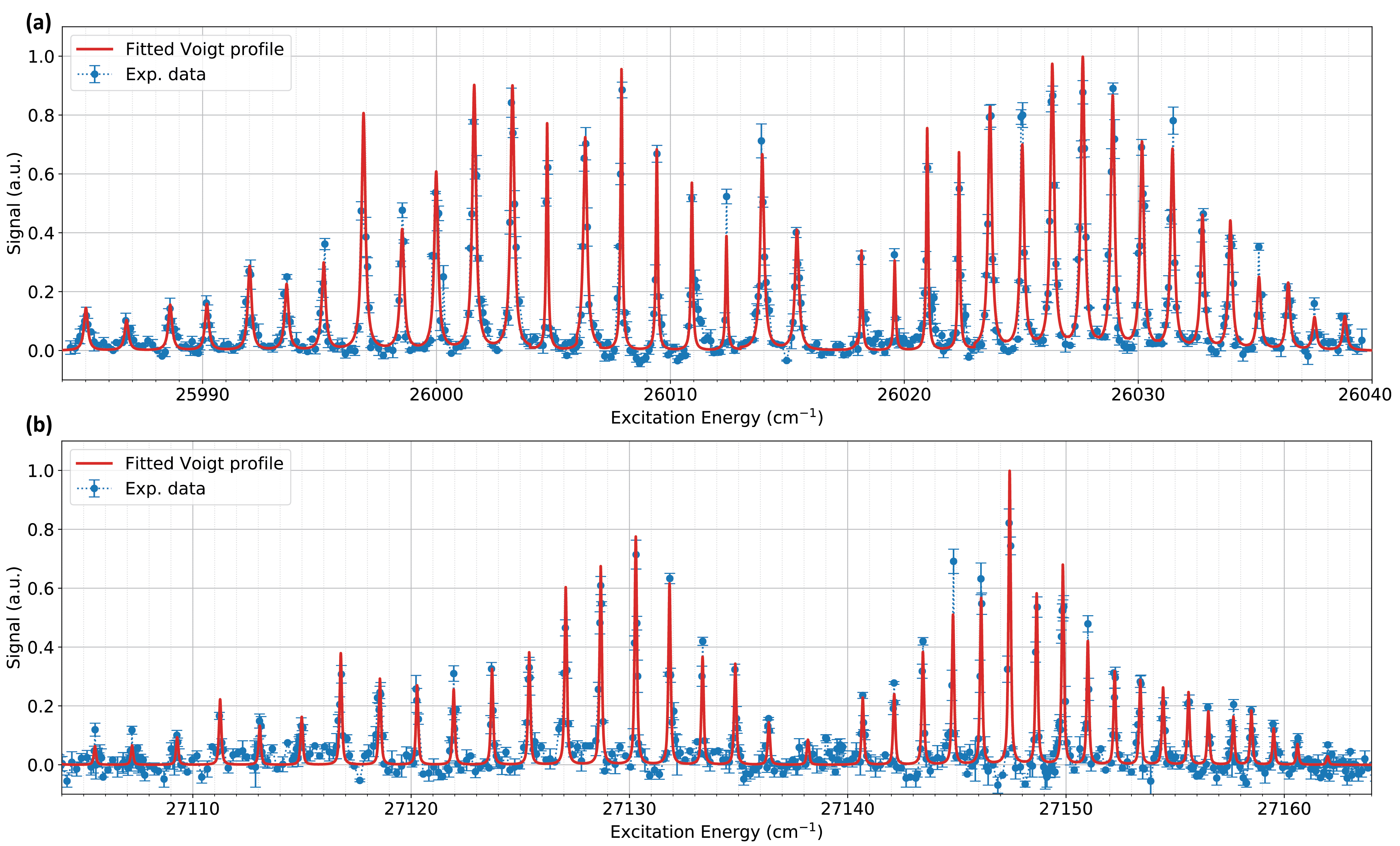}
    \caption{The experimental LIF data (blue dots) obtained by pulsed dye laser and the Voigt fittings (red curves) of \SiO{28} (a) (0,0) band and (b) (1,0) band of $B^2\Sigma^+ \leftarrow X^2\Sigma^+$
transition. }
    \label{PDLexp28}
\end{figure*}
While the main purpose of the pulsed dye laser experiment was to inform the subsequent higher-resolution measurement, analysis of the recorded data provided several useful results.
Fig.~\ref{28SiOFigure}(a-b) shows the LIF spectra along with a least-squares fitting of a Voigt profile to each rotational line, in which the Gaussian and Lorentzian contributions are estimated to be within the ranges of $0.5-0.8$~GHz (mostly corresponding to the Doppler broadening) and $3.0-6.0$~GHz (mainly corresponding to the laser linewidth), respectively.  
First, the measurements of the (0,0) and (1,0) bands in \SiO{28} provide a straightforward measurement of the separation of the first two vibrational states of the $B^2\Sigma^+$ vibrational  as $\omega_e^B - 2\omega_e x_e^B = 33657.29(13)$~GHz. 
Second, fitting the observed amplitudes of the rotational peaks indicates a rotational temperature of $T_r = 86(4)$~K. 
Given that the cell is held around 90~K, this indicated that molecular rotation is quickly thermalized after the molecules are created. 

\subsection{CW laser spectroscopy}

\begin{figure*}
    \centering
    \includegraphics[scale=0.9]{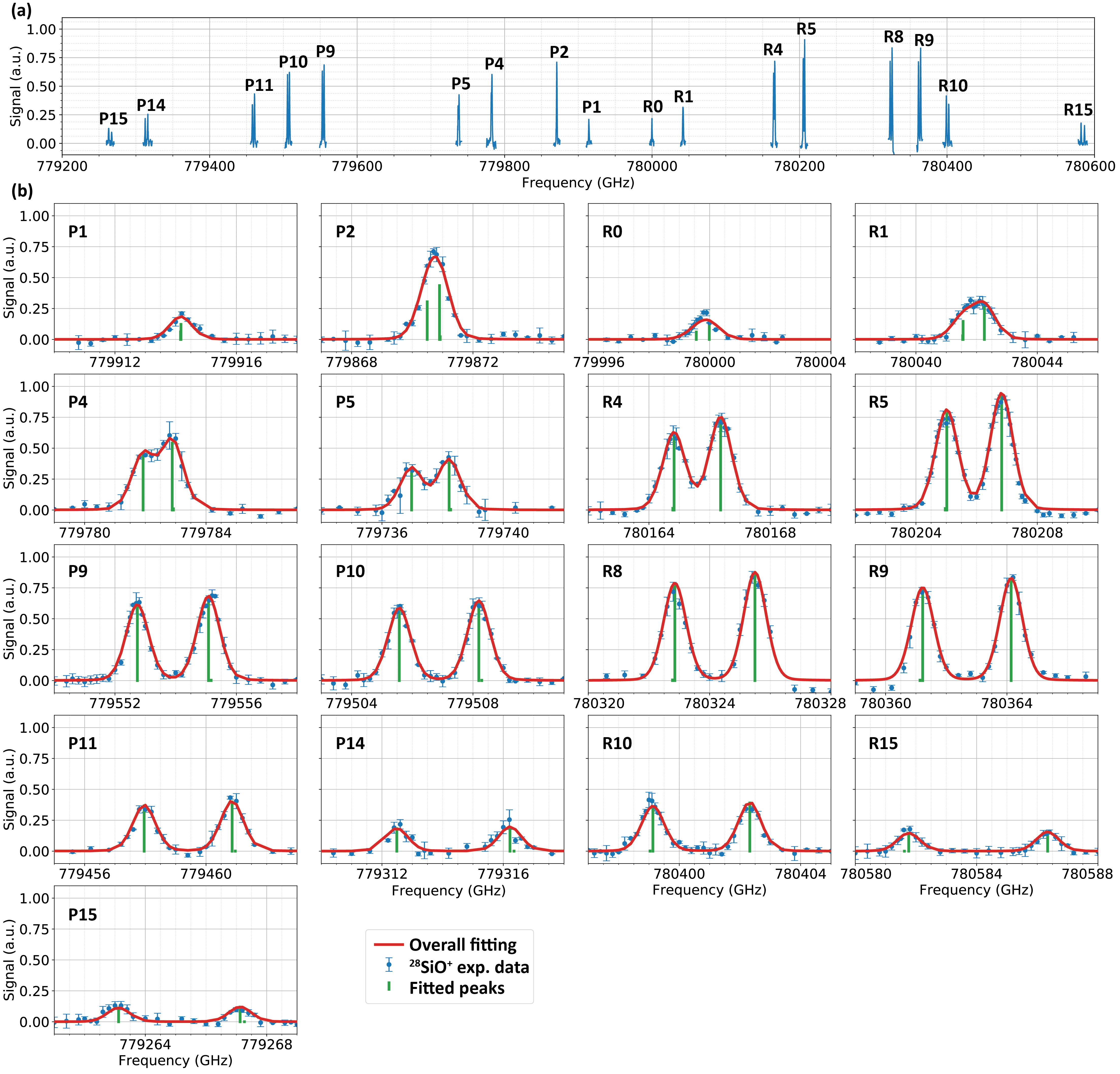}
    \caption{High-resolution LIF spectra of $B^2\Sigma^+ \leftarrow X^2\Sigma^+$ (0,0) band of \SiO{28}. (a) The overall spectrum of seventeen observed rotational lines, including nine P-branch and eight R-branch rotational lines (b) Seventeen spectra of single rotational lines and the corresponding fittings. The observed rotational lines are plotted by blue dots with error bars. The backgrounds are subtracted and the signal strengths are normalized to unity, which corresponds to the maximum signal strength detected in the experiment. The red curves are the overall fittings and the green sticks indicate the positions of each rotational transitions, which are calculated by substituting the fitted molecular constants into the theoretical model. The relative strengths of the peaks in each rotational line are calculated from the square of the transition dipole moment between the ground and excited states.}
    \label{28SiOFigure}
\end{figure*}

\begin{figure*}
    \centering
    \includegraphics[scale=0.9]{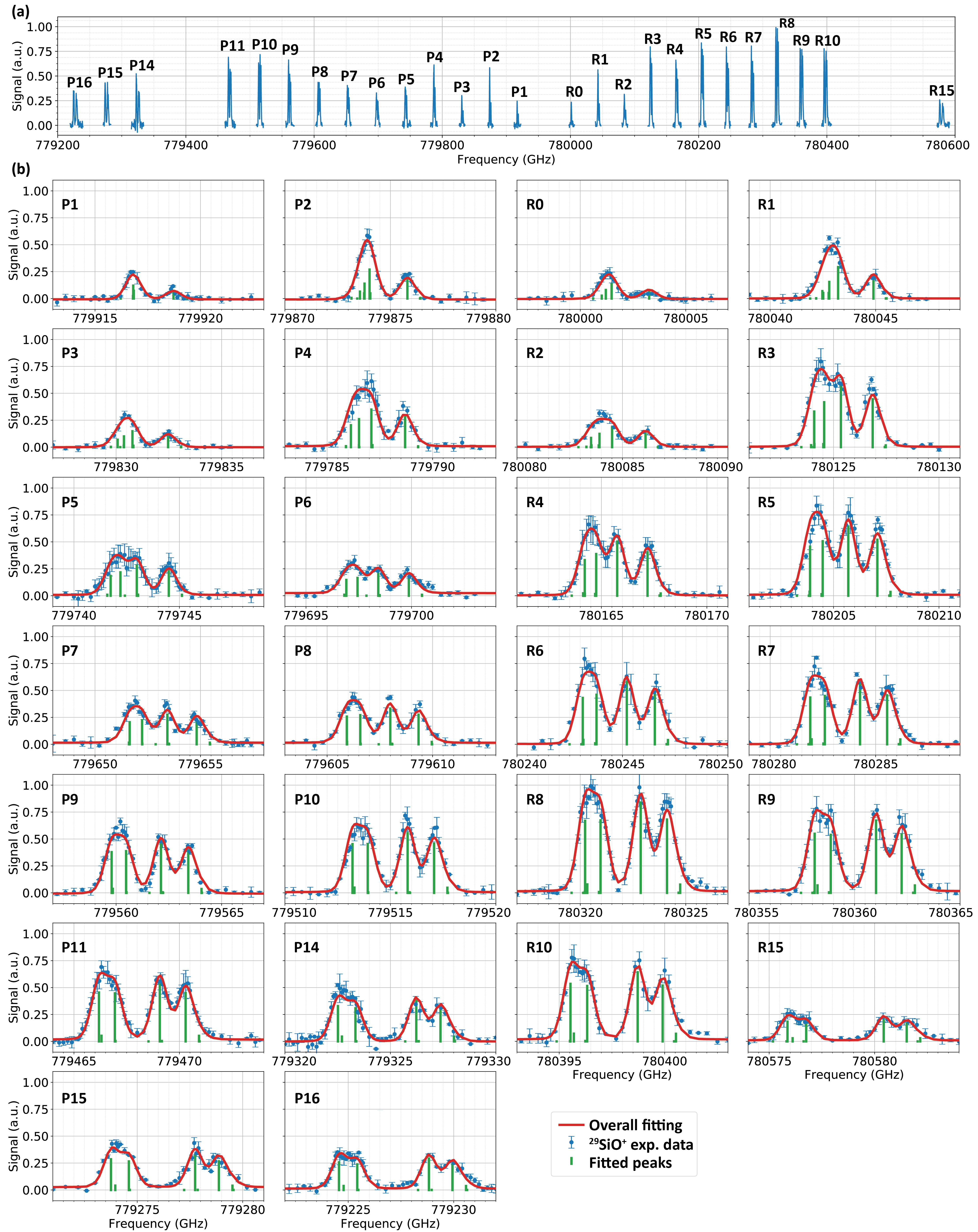}
    \caption{High-resolution LIF spectra of $B^2\Sigma^+ \leftarrow X^2\Sigma^+$ (0,0) band of \SiO{29}. (a) The overall spectrum of twenty-six observed  rotational lines, including fourteen P-branch and twelve R-branch lines. (b) Twenty-six spectra of single rotational lines and the respective fits.}
    \label{29SiOFigure}
\end{figure*}

The results of higher-resolution measurement are shown in Figs.~\ref{28SiOFigure} and \ref{29SiOFigure} for \SiO{28} and \SiO{29}, respectively.
The higher resolution scans required significantly more time, roughly half of a day per rotational line, and resulted in two changes to the experiment. 
First, the cryogenic cell temperature was raised to approximately 100~K to prevent O$_2$ ice from forming over the course of longer runs.
And second, long-term drifts in the ion signal, presumably due to changes in the ablation conditions, led to variations of the LIF amplitude from one rotational line to the next, preventing extraction of the molecular rotational temperature. As a result, in our fitting, we fix the relative strengths of the spin-rotational and hyperfine transitions \emph{within} a single rotational transition according to their respective transition moments, and allow a single fitted amplitude for each rotational transition. 


In this manner, custom software was created that diagonalizes
the relevant Hamiltonians, calculates the individual transition strengths between states, convolves those transition lines with a Gaussian lineshape of full width at half maximum (FWHM) of 840~MHz, which is consistent with the expected Doppler broadening of $\approx 800$~MHz, 
and fits to the data by minimizing $\chi^2$ with a gradient descent algorithm.
Once the best fit values were found, 68\% confidence intervals were found for each fitted parameter using the profiling method~\cite{cowan1998statistical}. 


\begin{table*}[ht!]
    \makegapedcells
    \centering
    \begin{tabular}{c|c c c|c c}
    \hline
    \hline
    \multirow{2}{*}{Parameters} &\multicolumn{3}{c|}{\SiO{28}} &\multicolumn{2}{c}{\SiO{29}}\\[1pt]\cline{2-6}
       & PDL & CW & Ref.& CW & Ref.\\
       \hline
        $T_0^B$ & 779962.32(5) & 779957.19(8) & $779948.27(6)^a$ 
        &779959.78(11)\\
        &&& $779957.30(12)^b$ &&\\
        $\omega_e^B-2\omega_ex_e^B$ & 33657.29(13)&  &33659.32(10)$^a$&\\
        $B_0^X$ & $21.513(4)^c$
        & 21.5137(9) & $21.51341(30)^a$ 
        &21.2435(12)& $21.2447(9)^f$\\
        &$21.555(4)^d$&& $21.5137(21)^b$ &&\\
        $B_0^B$ & 21.279(4)&21.2897(9)& $21.28796(30)^a$
        & 21.0217(12) & $21.0236(9)^f$\\
        &&&$21.2901(21)^b$&&\\
        $\alpha_e^B$& 0.123(5)&  &0.174139(14)$^a$&\\
        $D_0^X$ &  & $3.3(7)\times10^{-5}$ & $3.223(9)\times10^{-5~a}$
        & $3.2(7)\times10^{-5}$ & $3.2(7)\times10^{-5~f}$\\
        &&&$3.27(6)\times10^{-5~b} $&&\\
        $D_0^B$ &  & $3.5(7)\times10^{-5}$& $3.292(6)\times10^{-5~a}$
        & $3.2(8)\times10^{-5}$ & $3.4(7)\times10^{-5~f}$\\
        &&&$3.32(6)\times10^{-5~b} $&&\\
        $\gamma^X$&&0.009(18)& $0.006(25)^{e}$
        & 0.012(25) & $0.009(18)^f$\\
        &&&$0.12(12)^b$&&\\
        $\gamma^B$ &&0.286(18)& $0.287(24)^{e}$
        & 0.288(25) & $0.279(18)^f$\\
        &&&$0.420(12)^b$&&\\
        $b_F^X$ &&&&-0.60(22) & $-0.797(1)^g$\\
        $b_F^B$ &&&&-2.49(25)\\
        $t^X$ &&&&-0.04(30)& $-0.064(1)^g$\\
        $t^B$ &&&& 0.04(33)\\
    \hline
    \hline
    \multicolumn{5}{l}{\begin{tabular}{l}
    $^a$ Ref.~\cite{cameron1995fast1, cameron1995fast2}.\\[-5pt]
    $^b$ Ref.~\cite{ghosh1979confirmation}.\\[-5pt]
    $^c$ Our fitted results from  $B^2\Sigma^+ \leftarrow X^2\Sigma^+ (0,0)$ band. \\[-5pt]
    $^d$ Our fitted result from  $B^2\Sigma^+ \leftarrow X^2\Sigma^+ (1,0)$ band. \\[-5pt]
    $^{e}$  Extracted from (0,0) band data of Ref.~\cite{cameron1995fast1, cameron1995fast2}.\\[-5pt]
    $^f$ Predictions by scaling isotope constants of \SiO{28} measured in the CW experiment. \\[-5pt]
    $^g$ Ref.~\cite{knight1985generation}.
    \end{tabular}} 
    \end{tabular}
    \caption{\SiO{28} and \SiO{29} molecular constants (Unit: GHz). PDL denotes constants extracted from the low-resolution pulsed-dye laser measurement, while CW denotes constants extracted from the higher resolution continuous wave laser measurement. The constants in the CW column denote our recommended values. The exception to this are the values of $b_F^X$ and $t^X$, which are presumably better determined by a measurement in a solid neon matrix~\cite{knight1985generation}.}
    \label{SiOMolConst}
\end{table*}

The resulting fits for \SiO{28} are shown in Fig. \ref{28SiOFigure}. 
The fits yield a $\chi^2/DOF$ of 0.96 when assuming the data is Poissonian distributed, indicating a stable, well-behaved experiment and reasonable fit.
The fits also appear to satisfactorily reproduce the observed spectra.
The extracted molecular constants are shown and compared to previous results in Tab.~\ref{SiOMolConst}. 
The transition energy $T_0^B$, vibrational term $\omega_e^B-2\omega_ex_{e}^B$ and the rotational constants are in agreement with the previous results, however, the extracted spin-rotational constants $\gamma^X$ and $\gamma^B$ show some deviation from previously reported results.  

For $\gamma^X$, our result agrees with that reported by Ghosh et al.~\cite{ghosh1979confirmation}.
Cameron et al. \cite{cameron1995fast1, cameron1995fast2} and the later work of the same group \cite{rosner1998study}, also report $\gamma^X=0.34416(30)$~GHz, but 
their value comes from a deperturbation analysis that considers the perturbations in the X$^2\Sigma^+$ state from the nearby A$^2\Pi$ state.
Thus, their $\gamma^X$ cannot be directly compared to the one reported here. 
Therefore, we refitted the low-lying rotational transitions ($N < 40$) of the (0,0) band reported by Cameron et al. with our phenomenological model and found $\gamma^X = 0.006(25)$~GHz, which agrees with our reported value. 

Cameron et al.~\cite{cameron1995fast1,cameron1995fast2} also report data for $\gamma^B=0.2986(27)$~GHz.
When we refit their reported low-lying rotational transitions, in the same manner as we did for the X state, we find $\gamma^B = 0.287(24)$~GHz, which is compatible with our result.
The small change with refitting is presumably due to the lack of perturbation of the B state in their analysis. 
The value reported by Ghosh et al. is incompatible with our result, but those workers do not appear to use the relative line strength as employed here and are therefore sensitive only to $|\gamma^B-\gamma^X|$.
This leads to significant covariance in the two spin-rotation parameters and they therefore state that ``individual [spin-rotation parameters] cannot be assigned with any certainty."


Fig.~\ref{29SiOFigure} shows a high-resolution specrum of \SiO{29}. 
The data are fitted in the same way as the \SiO{28} with the addition of the relevant hyperfine Hamiltonians for each state. 
If the data are assumed to be Poissionian distributed, the best fit value yields a $\chi^2/DOF = 1.34$ 
, which for the current number of degrees of freedom (1366)
must be rejected. 
As the resulting fit reproduces the data reasonably well, we conclude that data is likely super-Poissonian, presumably due to extra fluctuations introduced by ablation of the pressed $^{29}$Si target, and increase the standard deviation in our fits such that the $\chi^2/DOF = 1$.
While this procedure does not change the best fit values, it does increase the confidence intervals on the reported parameters. 


The extracted molecular constants for \SiO{29} are listed in the second-to-last column of Tab. \ref{SiOMolConst}. 
Comparing to the molecular constants of \SiO{28}, the \SiO{29} transition energy $T_0^B$ has an isotope shift of $2.59(14)$~GHz, which is due to the shift in $T_e$ and the zero-point energy
Further, the rotational constants are a few percent smaller due to the greater reduced mass. 
The spin-rotation constants are indistinguishable from those of \SiO{28} at the current experimental resolution.
By using the relative transition strength in the fitting procedure we are able to resolve the Fermi contact parameter for both the X and B states to be $-0.60(22)$~GHz and $-2.49(25)$~GHz, respectively. 
The larger Fermi contact interaction in the B state is consistent with the molecular orbital description provided in Refs.~\cite{stollenwerk2017electronic,cai1998ab}, where the $B^2\Sigma^+ \leftarrow X^2\Sigma$ transition involves electron promotion to an orbital with a larger Si contribution.
The dipolar hyperfine parameters for both states are merely constrained by our fit. 

While there has been no previous measurement of the hyperfine structure of the \SiO{29} B$^2\Sigma^+$ state to our knowledge, Knight et al.~\cite{knight1985generation} report measurements of the X$^2\Sigma^+$ state hyperfine parameters in a neon matrix.
These values are in agreement with our measured values. 
A review of the literature to compare hyperfine parameters determined in a neon matrix to those measured in gas phase suggests that the values are typically within a few percent of one another \cite{aldegunde2018hyperfine}.
Therefore, the more accurate values of Knight et al. \cite{knight1985generation} likely provide the best estimate of the \SiO{29} ground state hyperfine structure. 
If we fix $b_F^X$ and $t^X$ to the Knight et al. values \cite{knight1985generation}, our estimates of the B state hyperfine parameters become $b_F^B = -2.72(21)$~GHz and $t^B = -0.03(32)$~GHz. 




\subsection{Lifetime of the B$^2\Sigma$\+ state}
\begin{figure}
    \centering
    \includegraphics[scale=0.9]{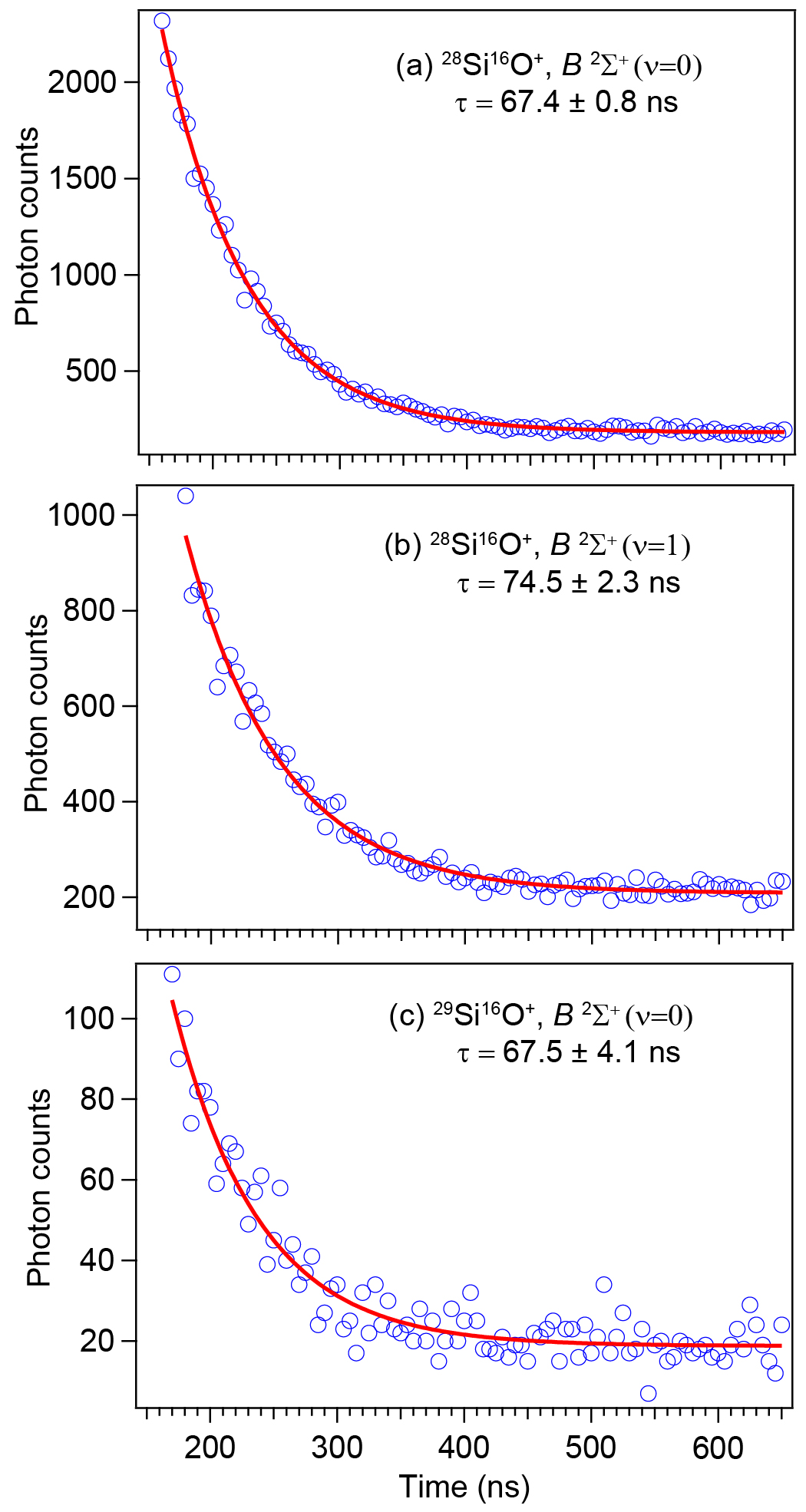}
    \caption{Fluorescence decay curves (blue circles) and exponential fit (red traces) from (a) \SiO{28} $B^2\Sigma^+ (v=0)$, (b) \SiO{28} $B^2\Sigma^+ (v=1)$ and (c) \SiO{29} $B^2\Sigma^+ (v=0)$. The lifetimes are determined to be 67.4 (0.8) ns, 74.5 (2.3) ns and 67.5 (4.1) ns, respectively.}
    \label{fig:lifetime}
\end{figure}

The recorded LIF data with pulsed dye laser excitation can also be fitted by an exponential decay function to determine the spontaneous emission lifetime of the excited  $B^2\Sigma^+$ states, as shown in Fig.~\ref{fig:lifetime}. 
From this data, we observe the lifetime of $B^2\Sigma^+ (v=0)$ of \SiO{28} to be 67.4(0.8) ns, which is consistent with the previously reported value of 66(2) ns \cite{stollenwerk2017electronic}, but in tension with the value of 69.5(0.6) ns reported by Ref.~\cite{scholl1995beam}. 
For the first-excited vibrational level $v=1$ we observe a longer lifetime of 74.5(2.3) ns, which is in good agreement with the previous value of 72.4 (0.5) ns ~\cite{scholl1995beam}.
The longer lifetime of the excited vibrational level is due to a decrease in the transition dipole moment with the increased internuclear distance, as shown in the theoretical calculations~\cite{li2019laser,cai1998internally,Chattopadhyaya2003ElectronicSO}. 
The same lifetime trend is also observed in CO$^+$, where it is due to mixing of a low-lying electronic configuration (...$4\sigma^2 1\pi^3 5 \sigma^1 2\pi^1$) with the B$^2\Sigma$\+ state configuration of ...$4\sigma^1 1\pi^4 5 \sigma^2$ with increasing internuclear distance~\cite{scholl1995beam,marian1989theoretical}.
Finally, the lifetime of $B^2\Sigma^+ (v=0)$ of \SiO{29} is determined to be 67.5(41) ns. 

\section{Systematics}
The error bars reported here are statistical standard errors.
Systematic shifts, which could include, e.g., Zeeman and Stark shifts, non-uniform Doppler shifts, and laser wavelength calibration are possible.
An order of magnitude estimate of the systematic error can be made by the following considerations. 
Given typical laboratory stray fields, Zeeman and Stark shifts on the order of a few MHz are possible. 
The ablation process could lead to a non-uniform gas flow and therefore a Doppler shift of the observed lines. 
Analysis of the residuals of a typical lineshape suggests that such a shift is likely $< 5$~MHz. 
Finally, the wavemeter reported inaccuracies are $\approx 2$~MHz and the laser frequency fluctuation is within $\sim 10$~MHz. 
From the covariance matrix resulting from the aforementioned fit, we estimate that fluctuations of this order would likely lead to corrections to $B_0$, $D_0$, $\gamma$, and $b_F$ of roughly $< 50$~kHz, $< 0.25$~kHz, $< 0.5$~MHz, and $< 20$~MHz, respectively. 

\section{Quantum logic with \SiO{29}}
\begin{figure}
    \centering
    \includegraphics[scale=0.9]{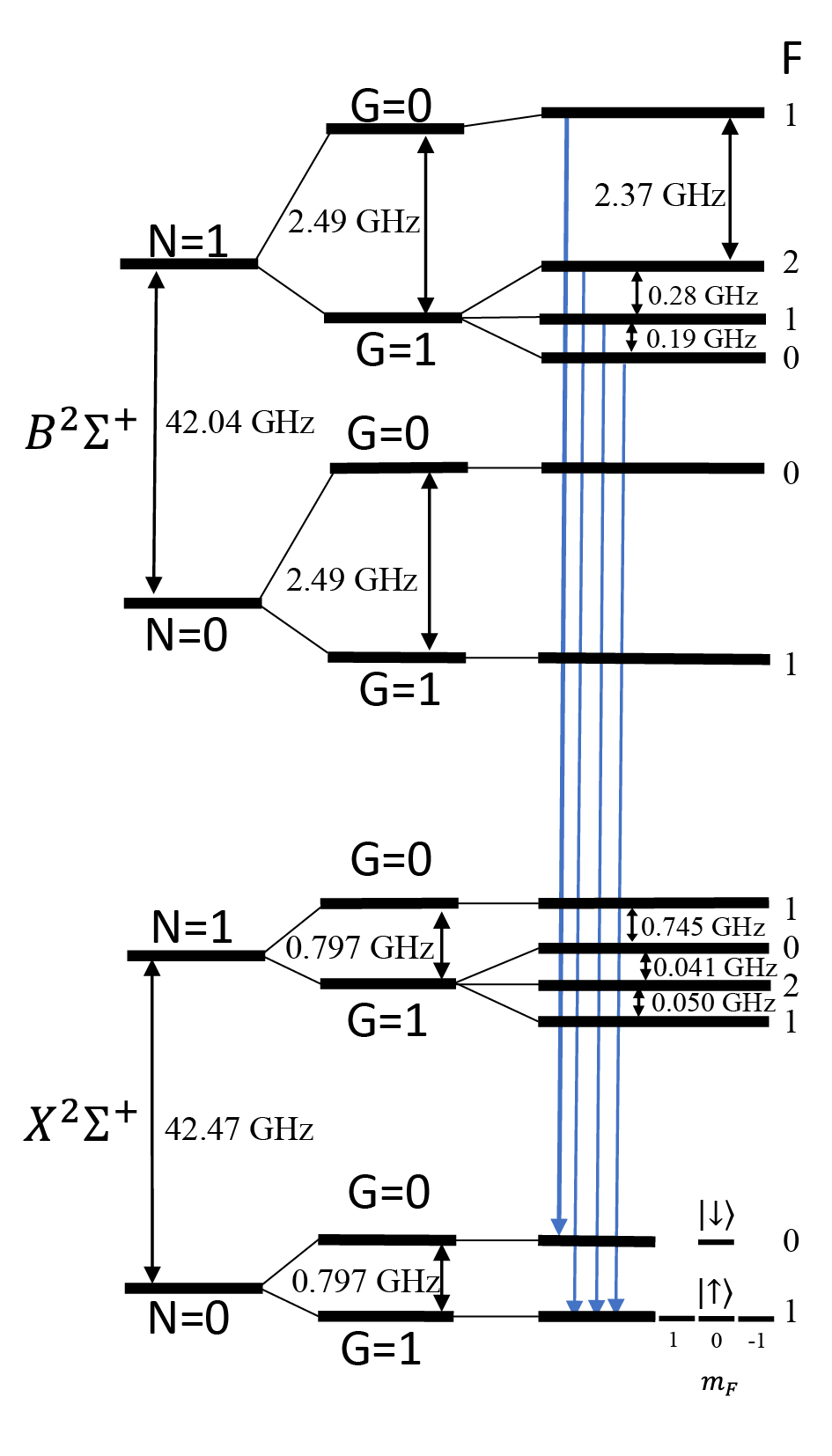}
    \caption{Low-lying structure of the $B^2\Sigma^+$ and $X^2\Sigma^+$ states of \SiO{29}. Blue arrows show the strong $\Delta G = 0$ electric dipole transitions. $\Delta G = 1$ transitions are allowed, but are significantly weaker due to the small spin-rotation interaction (relative to the hyperfine interaction). The qubit states are defined as $\ket{\downarrow}\equiv\ket{X,N=0,G=0,F=0}$ and $\ket{\uparrow}\equiv\ket{X,N=0,F=1,m_F=0}$. 
    }
    \label{fig:29SiOstructure}
\end{figure}

\begin{figure}
    \centering
    \includegraphics[scale=0.82]{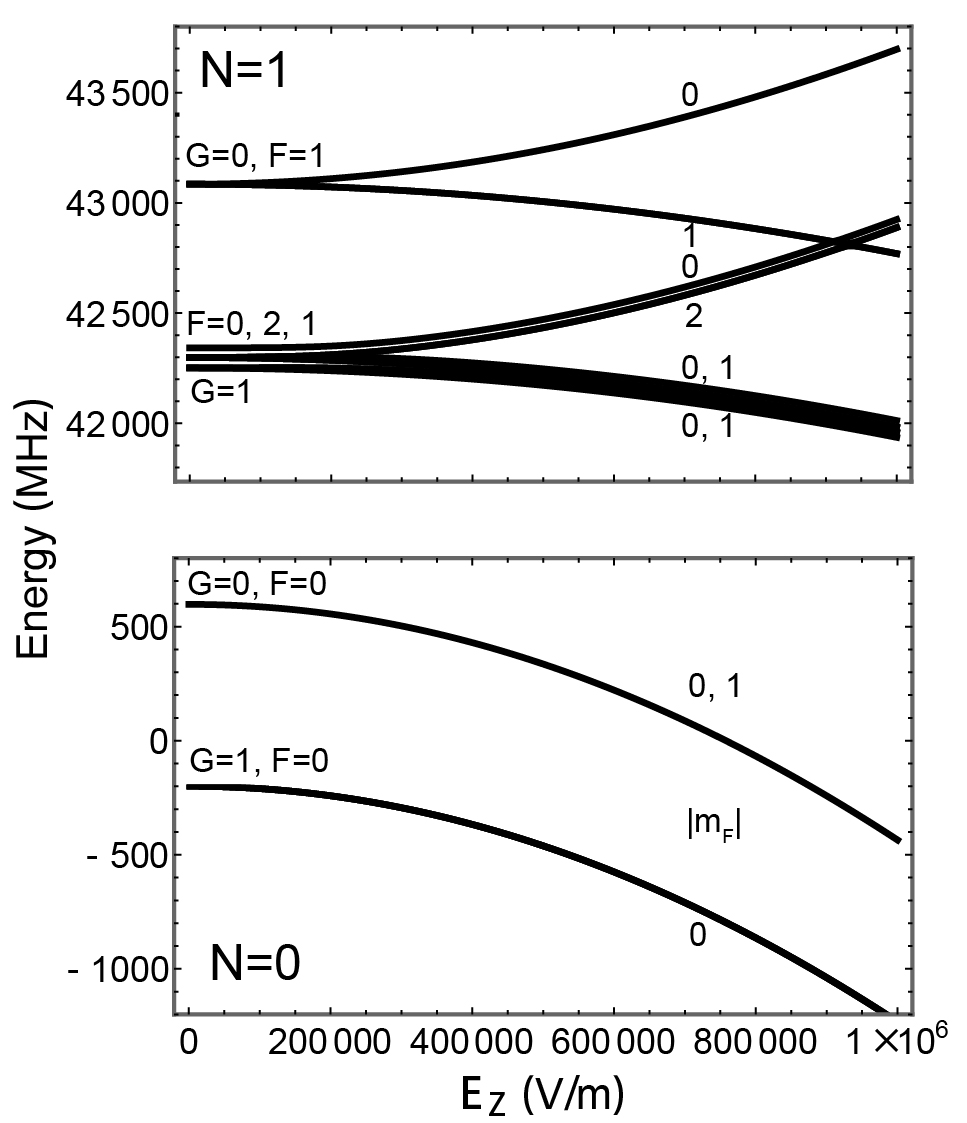}
    \caption{Calculated DC Stark shift of the $X{}^2\Sigma^+$ state of \SiO{29} based on the spectroscopically determined structure parameters.  The qubit is first-order insensitive to electric and magnetic field noise, allowing long coherence times, even in an ion trap environment.}
    \label{29SiOstarkshift}
\end{figure}


With the presented spectroscopic data, the relevant structure for using \SiO{29} in quantum logic operations is determined and shown in Fig.~\ref{fig:29SiOstructure}. 
Each rotational state in the $X$ and $B$ electronic states is split into two hyperfine states $G=0, 1$, and the $G=1$ state is split into three states with different total angular momentum, $F$, when $N>0$.  

The presence of $F = 0$ states provides convenient means for qubit state-preparation and measurement.
Namely, state preparation can be accomplished by first rotationally cooling the molecule to the $N=0$ state~\cite{stollenwerk2020cooling}, followed by optical pumping on the weakly-allowed $\ket{B,N=1, G=1, F=1} \leftarrow \ket{X,N=0,G=0,F=1}$ transition, while applying rotational and vibrational rempumping.
Because $\Delta G \neq 0$ transitions are weak, the $\ket{B,N=1,G=0,F=1}$ state decays to the the $\ket{\downarrow}\equiv\ket{X,N=0,G=0,F=0}$ with 99.994\% probability. 
In this manner, the molecular qubit can be initialized in the $\ket{\downarrow}$ state with high fidelity after a single scattering event. 
Defining the other state of the qubit as $\ket{\uparrow} \equiv \ket{X,N=0,F=1,m_F=0}$, preparation of an arbitrary qubit state can be realized by driving a microwave M1 or optically stimulated Raman transition resonant on the $\approx 800$~MHz hyperfine transition frequency.

Measurement of the qubit state can be accomplished by applying a laser on resonance with the $\ket{B,N=1,G=1, F=0} \leftarrow \ket{X,N=0,G=1, F=1}$ transition, while applying the appropriate vibrational and rotational repumping.
Because spontaneous emission between two $F = 0$ levels is strictly forbidden by angular momentum conservation, this will lead to scattering of photons only if the qubit is found in $\ket{\uparrow}$.
Detecting any resulting fluorescence, thus allows the determination of the qubit state. 
Alternatively, state detection could also be performed by using laserless electric field gradient gates and monitoring a co-trapped atomic ion~\cite{hudson2020laserless}, or pumping the $\ket{\uparrow}$ state to the $\ket{B,N=0,G=1, F=0}$ state via a microwave transition $\ket{X, N=1} \leftrightarrow \ket{X, N=0}$ and detecting fluorescence on the P1 transition $\ket{B, N=0} \leftrightarrow \ket{X, N=1}$. 


Single qubit gates can be accomplished with microwave radiation at the $\approx 800$~MHz qubit frequency or via a laser-driven Raman transition.

Two-qubit gates can be realized in a number ways. 
The laser-based Raman gates employed in atomic ion quantum computing, e.g. the M\o{}lmer-S\o{}rensen gate~\cite{sorensen1999quantum,vitanov2017stimulated}, 
can be applied to molecules without modification. While these gates are well understood, they require near motional ground state cooling and stablization of the relative phase between two laser frequency. 
A potentially less technologically demanding alternative is to use the recently proposed electric field gradient gates (EGGs) \cite{hudson2020laserless}
, which requires only the application of microwave gradient electric field that is near resonance with the $\ket{X,N=1}\leftarrow\ket{X,N=0}$ rotational transition at $\approx 42.5$~GHz.

The EGGs interaction proceeds by apply an oscillating quadrupole electric field via the ion trap itself, which results in a position-dependent electric field that couples the internal state of the qubit to the ion motion. 
This can be used to entangle two trapped \SiO{29} ions as follows. 
For two trapped \SiO{29} ions, microwave radiation is used to transfer population from $\ket{\uparrow \uparrow} \rightarrow \ket{ee}$, where $\ket{e}$ is some excited state that also couples to $\ket{\downarrow}$. 
This excited state is necessary since the $\ket{\uparrow} \leftrightarrow \ket{\downarrow}$ transition is electric dipole forbidden.
For \SiO{29}, we choose $\ket{e}$ to be $\ket{X, N = 1, G = 0}$ due to its strong coupling to the $\ket{\uparrow}$ state.
A M\o lmer-S\o rensen gate can then be applied via the EGGs interaction on the $\ket{e} \leftrightarrow  \ket{\downarrow}$ transition, producing $\ket{\psi} = \frac{1}{\sqrt{2}} \left( \ket{ee} + i \ket{\downarrow \downarrow} \right)$. 
The weak coupling  between the $\ket{e}$ and $\ket{\downarrow}$ states can be remedied by simply increasing the voltage applied to the rods since the M\o lmer-S\o rensen gate is highly sensitive to detuning errors.
Finally, microwave radiation can be used to produce  $\ket{\psi} = \frac{1}{\sqrt{2}} \left( \ket{\uparrow \uparrow + i \ket{\downarrow \downarrow}} \right)$.
Simulations show that this sequence can reasonably yield Bell-state fidelities $\geq 0.99$ with gate times on the order of a few milliseconds.




Decoherence of $\{\downarrow,\uparrow\}$ qubits can be anticipated from a variety of sources, including blackbody radiation, collisions with background gas, and Zeeman and Stark shifts.
Ref.~\cite{stollenwerk2020cooling} observes that \SiO{28} prepared in the $N = 0$ state is lost with a time constant of tens of seconds and suggests this may be due to a combination of blackbody radiation and background gas collisions.
As the qubit states are zero-field clock states, magnetically-limited coherence times in excess of a second are expected~\cite{Christensen2020high}.
Finally, with the structure of \SiO{29} determined, the Stark shift may be calculated as shown in Fig.~\ref{29SiOstarkshift}.
As expected for a Hund's (b)$_{\beta S}$ molecule~\cite{hudson2018dipolar}, the chosen qubit states exhibit a small differential Stark shift of $\approx 6.8 \times 10^{-7}E^2$~[Hz/(V/m$^2$)].
Thus, in total qubit coherence time in excess of seconds should be expected.

\section{Summary}
We have performed laser-induced fluorescence spectroscopy of the $B^2\Sigma^+\leftarrow X^2\Sigma^+$ transition in both \SiO{28} and \SiO{29}.
The ions were produced by reaction of Si\+ with O$_2$ in a cryogenic cell at roughly 100~K. 
Low-resolution spectra of the (0,0) and (1,0) bands of \SiO{28} and the (0,0) band of \SiO{29} were obtained with a pulsed dye laser ($\approx 1.2$~GHz linewidth), yielding molecular constants of both states, the lifetimes of $B^2\Sigma^+ (v=0,1)$ states of \SiO{28}, and the lifetime of the $B^2\Sigma^+ (v=0)$ of \SiO{29}.  
Using a narrowband cw Ti:sapphire laser ($\approx10$~MHz linewidth), we have observed seventeen and twenty-six high-resolution rotational lines of the (0,0) band of \SiO{28} and \SiO{29}, respectively. 
These data are used to extract more precise molecular constants for both ions, including the determination of the Fermi contact interaction constant for both the $B^2\Sigma^+$ and $X^2\Sigma^+$ states in the gas phase.
The observed spin-rotation constants show some disagreement with two previous, conflicting reports.
We show this disagreement is due to a lack of experimental resolution in one case and a differing model in the other. 
Finally, we outline how the determined and fortuitous structure of \SiO{29} can be used to perform state preparation and measurement of a qubit and one/two-qubit gates, as well as discuss several potential sources of decoherence.

\section{Acknowledgements}
This work was supported by the U.S. Department of Energy, Office of Science, Basic Energy Sciences, under award no. DE- SC0019245.
The authors thanks Ivan Antonov, Micheal Heaven, Brian Odom, and Tim Steimle for helpful comments on the manuscript. 


\bibliographystyle{elsarticle-num} 
\bibliography{cas-refs}





\end{document}